\documentclass[preprint]{aastex}
\usepackage{ulem}
\usepackage{color}
\usepackage{cancel}
\usepackage{pdflscape}

\newcommand{\hCn}{H$^{13}$CN}
\newcommand{\hcN}{HC$^{15}$N}
\newcommand{\hCN}{H$^{13}$C$^{15}$N}
\newcommand{\cratio}{$^{12}$C/$^{13}$C}
\newcommand{\nratio}{$^{14}$N/$^{15}$N}
\newcommand{\kms}{km\,s$^{-1}$}

\begin{document}
\slugcomment{Accepted to ApJ Letters, 2016/05/30}

\title{ALMA Observations of HCN and its Isotopologues on Titan}


\author{Edward M. Molter$^{1,2}$, C. A. Nixon\footnotemark[1], M. A. Cordiner$^{1,2}$, J. Serigano\footnotemark[3], P. G. J. Irwin\footnotemark[4], N. A. Teanby\footnotemark[5], S. B. Charnley\footnotemark[1], J. E. Lindberg\footnotemark[1]}
\email{edward.m.molter@nasa.gov}

\footnotetext[1]{NASA Goddard Space Flight Center, 8800 Greenbelt Road, Greenbelt, MD 20771, USA}
\footnotetext[2]{Department of Physics, Catholic University of America, Washington, DC 20064, USA}
\footnotetext[3]{Department of Earth and Planetary Sciences, Johns Hopkins University, Baltimore, MD 21218, USA}
\footnotetext[4]{Atmospheric, Oceanic, and Planetary Physics, Clarendon Laboratory, University of Oxford, Parks Road, Oxford, OX1 3PU, UK}
\footnotetext[5]{School of Earth Sciences, University of Bristol, Wills Memorial Building, Queens Road, Bristol, BS8 1RJ, UK}
\begin{abstract}

We present sub-millimeter spectra of HCN isotopologues on Titan, derived from publicly available ALMA flux calibration observations of Titan taken in early 2014. We report the detection of a new HCN isotopologue on Titan, \hCN{}, and confirm an earlier report of detection of DCN. We model high signal-to-noise observations of HCN, \hCn{}, \hcN{}, DCN, and \hCN{} to derive abundances and infer the following isotopic ratios: \cratio{} = 89.8 $\pm$ 2.8, \nratio{} = 72.3 $\pm$ 2.2, D/H = (2.5 $\pm$ 0.2)$\times$10$^{-4}$, and HCN/\hCN{} = 5800 $\pm$ 270 (1$\sigma$ errors). The carbon and nitrogen ratios are consistent with and improve on the precision of previous results, confirming a factor of $\sim$2.3 elevation in \nratio{} in HCN compared to N$_2$ and a lack of fractionation in \cratio{} from the protosolar value. This is the first published measurement of D/H in a nitrile species on Titan, and we find evidence for a factor of $\sim$2 deuterium enrichment in hydrogen cyanide compared to methane. The isotopic ratios we derive may be used as constraints for future models to better understand the fractionation processes occurring in Titan's atmosphere.

\end{abstract}


\section{Introduction}
\label{Intro}

Titan's thick (1.45 bar) atmosphere is primarily composed of molecular nitrogen (N$_2$, $\sim$98\%) and methane (CH$_4$, $\sim$1.5\%), but also hosts a myriad of trace organic species \citep[for a recent review, see][]{bezard14}.  Titan's complex photochemistry is born from the photodissociation of methane and nitrogen in the upper atmosphere by radiaton and charged particle impacts. The resulting ions recombine into simple hydrocarbons and nitriles \citep[e.g.,][]{wilson04}, which react further to produce more complex organic molecules and eventually agglomerate to become the grains that form Titan's haze layers. The abundance of such a rich organic chemistry as well as the presence of a liquid solvent on the moon's surface \citep[][]{stofan07} has motivated speculation that conditions on Titan may be suitable for biology \citep[e.g., ][]{khare86,sagan92,stevenson15}.  Understanding the abundances, distributions, and variability of photochemical products is essential to modeling the global circulation and chemistry of Titan's atmosphere. Isotopic ratios are a useful probe of the processes governing the physical and chemical evolution not only of Titan but of the solar system as a whole, documenting the history of each element and molecule from the proto-solar nebula to the planetary system we see today.

The most abundant nitrogen-bearing photodissociation product in Titan's atmosphere is hydrogen cyanide (HCN), which is formed principally through the sequence N + CH$_3$ $\rightarrow$ H$_2$CN + H $\rightarrow$ HCN + H$_2$ as well as many secondary processes \citep[][and references therein]{loison15}. The molecule has been well studied by previous ground-based and satellite observations. An infrared limb spectrum taken by the Voyager I spacecraft produced the first vertical abundance profile of the gas, with a vertical resolution of $\sim$200 km \citep[][]{coustenis91}. \citet{hidayat97} and \citet{marten02} used single-dish submillimeter observations of rotational transitions of HCN, \hCn{}, and \hcN{} from the IRAM 30 m telescope and the JCMT on Mauna Kea to produce disk-averaged HCN vertical profiles and determine the \cratio{} and \nratio{} ratios. Subsequent submillimeter observations by the Submillimeter Array \citep[][]{gurwell04} and the Herschel Space Observatory \citep[][]{courtin11} confirmed and refined these measurements. The arrival of the Cassini spacecraft to the Saturnian system in 2004 permitted the detection of infrared spectral lines of \hCn{} and \hcN{} using the CIRS instrument \citep[][]{vinatier07}, as well as extensive mapping of the vertical and horizontal distributions of HCN \citep{teanby07, vinatier10, koskinen11}.

The advent of ALMA provides the opportunity to probe Titan at submillimeter wavelengths with unprecendented sensitivity and spatial resolution. Since ALMA often uses Titan as a flux calibration source, a wealth of observations of the moon covering different parts of the submillimeter spectrum are available in the ALMA Science Archive.  These calibration observations can be used to generate significant science return despite relatively short integration times of around three minutes each 
\citep[e.g.][]{cordiner14,cordiner15,serigano16}.
In this paper we make use of several such observations to detect and model isotopes of HCN on Titan. 

\section{Observations and Data Processing}
\label{sec2}

We downloaded publicly available ALMA datasets taken between 3 April and 8 July 2014 that used Titan as a flux calibration source. This time period corresponds to less than 1\% of a Titan year, so we assume in our analysis that seasonal temperature and gas abundance changes are negligible. The data processing procedure we used was very similar to that of \citet{cordiner15}. Each dataset was flagged and calibrated by the North American ALMA Science Center using the standard data reduction procedures contained in the NRAO's CASA software version 4.5.0. The observed continuum level was scaled to match the Butler-JPL-Horizons 2012 flux model, which is expected to be accurate to within 15\% (see ALMA Memo \#594). Imaging was performed using standard CASA routines. Deconvolution of the ALMA point-spread function was performed using the Hogbom algorithm with natural visibility weighting. Details of each observation are shown in Table \ref{obsDetails}.

A disk-averaged spectrum was extracted from each data cube by integrating each channel within a circular region around the center of Titan encompassing all connected pixels for which the moon's emission was observed above the 3$\sigma$ noise level. The flux outside this region was found to be negligible.  Each spectrum was Doppler corrected to Titan's rest frame and converted to distance-independent radiance units using the distance and radial velocity of Titan with respect to the observer given by JPL Horizons\footnotemark[6] (see Table \ref{obsDetails}). The observed spectra are shown in Figure \ref{spectra_hcn}.

\footnotetext[6]{http://ssd.jpl.nasa.gov/horizons.cgi}

\section{Spectral Line Modeling and Results}
\label{sec3}

The model spectra were calculated using the line-by-line radiative transfer module of the NEMESIS atmospheric retrieval code \citep[][]{irwin08}.  Spectral line wavenumbers and intensities were taken from \citet{ahrens02}, \citet{brunken04}, and \citet{fuchs04} as recommended by the Cologne Database for Molecular Spectroscopy \citep[CDMS; ][]{muller01} and converted into HIgh-resolution TRANsmission molecular absorption database (HITRAN) 2004 format \citep[][]{rothman05} following the procedures described in the HITRAN online documentation\footnotemark[7]. The Lorentzian broadening half-width at 296 K ($\gamma$) was assumed to be 0.13 \citep[][]{yang08} after correcting for N$_2$ broadening as in \citet{teanby10}, with a temperature-dependence exponent ($\tau$) of 0.75 \citep[][]{devi04} as recommended by HITRAN. Partition functions were derived for each isotopologue using a third order polynomial fit of the partition function data provided by CDMS. The reference atmosphere and collision-induced absorption parameters used in this paper are the same as in \citet{teanby13} except that here the atmosphere is allowed to extend to 1200 km above Titan's surface.

\footnotetext[7]{http://hitran.org/docs/jpl-cdms-conversion}

Accurate modeling of a disk-averaged spectrum around Titan requires accounting for limb brightening due to the moon's extended atmosphere. We follow the method described in the Appendix of \citet{teanby13}, which prescribes calculating a weighted sum of discrete spectral radiances at different radii from Titan's center.  Seventy-two averaging points are sufficient to accurately model the spectrum for the strongest observed lines.

The continuum emission from Titan modeled by NEMESIS is $\sim$3\% less than the continuum level of the data in every spectral region we analyze.  We presume this discrepancy is caused by a slight difference between the NEMESIS model and the Butler-JPL-Horizons 2012 flux model used by the NRAO to self-calibrate Titan.  Since the offset is the same across all wavelengths the data is simply multiplied by a constant factor such that the continuum level matches the model in a line-free spectral window.

\begin{figure*}
\includegraphics[width=0.33\textwidth]{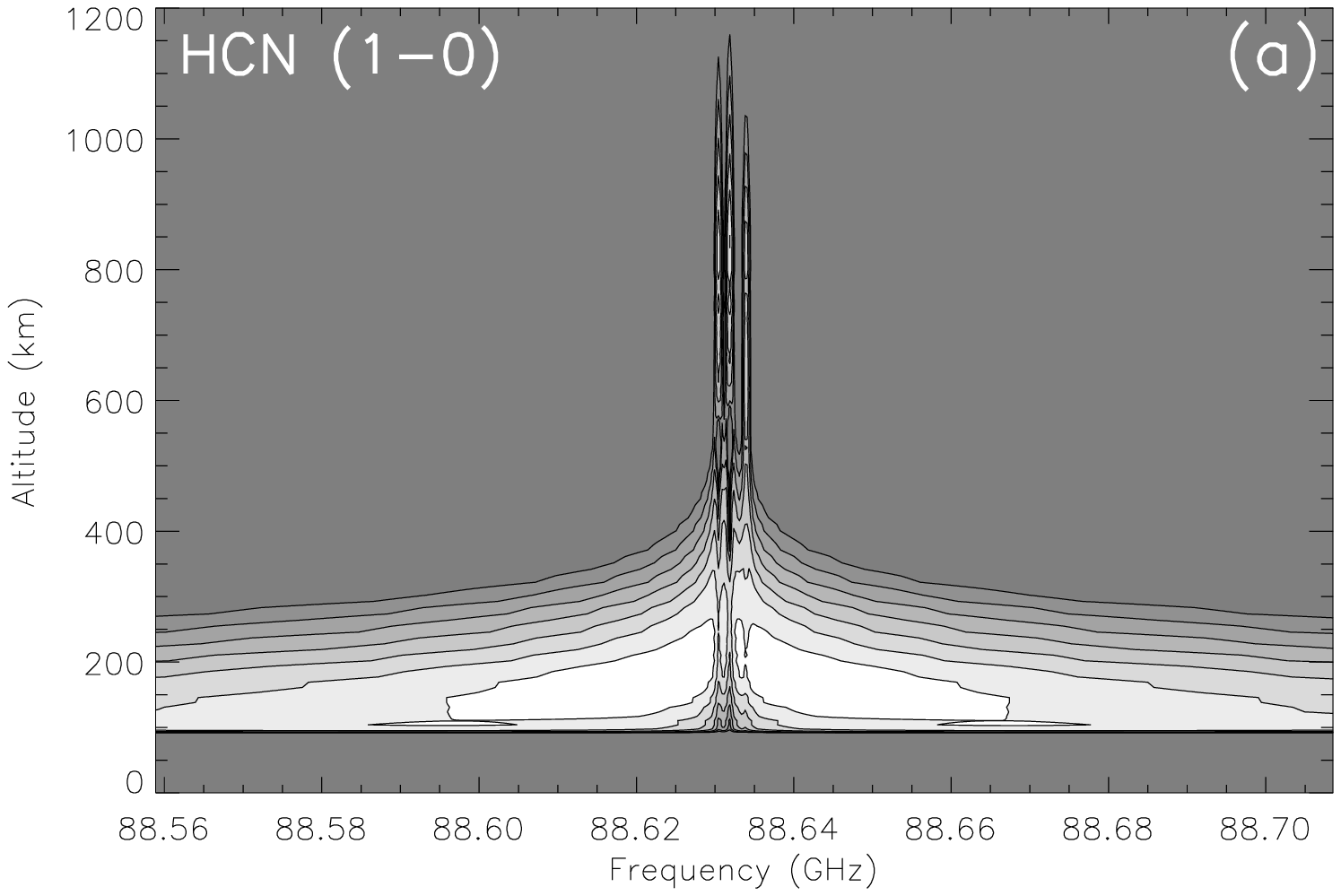}
\includegraphics[width=0.33\textwidth]{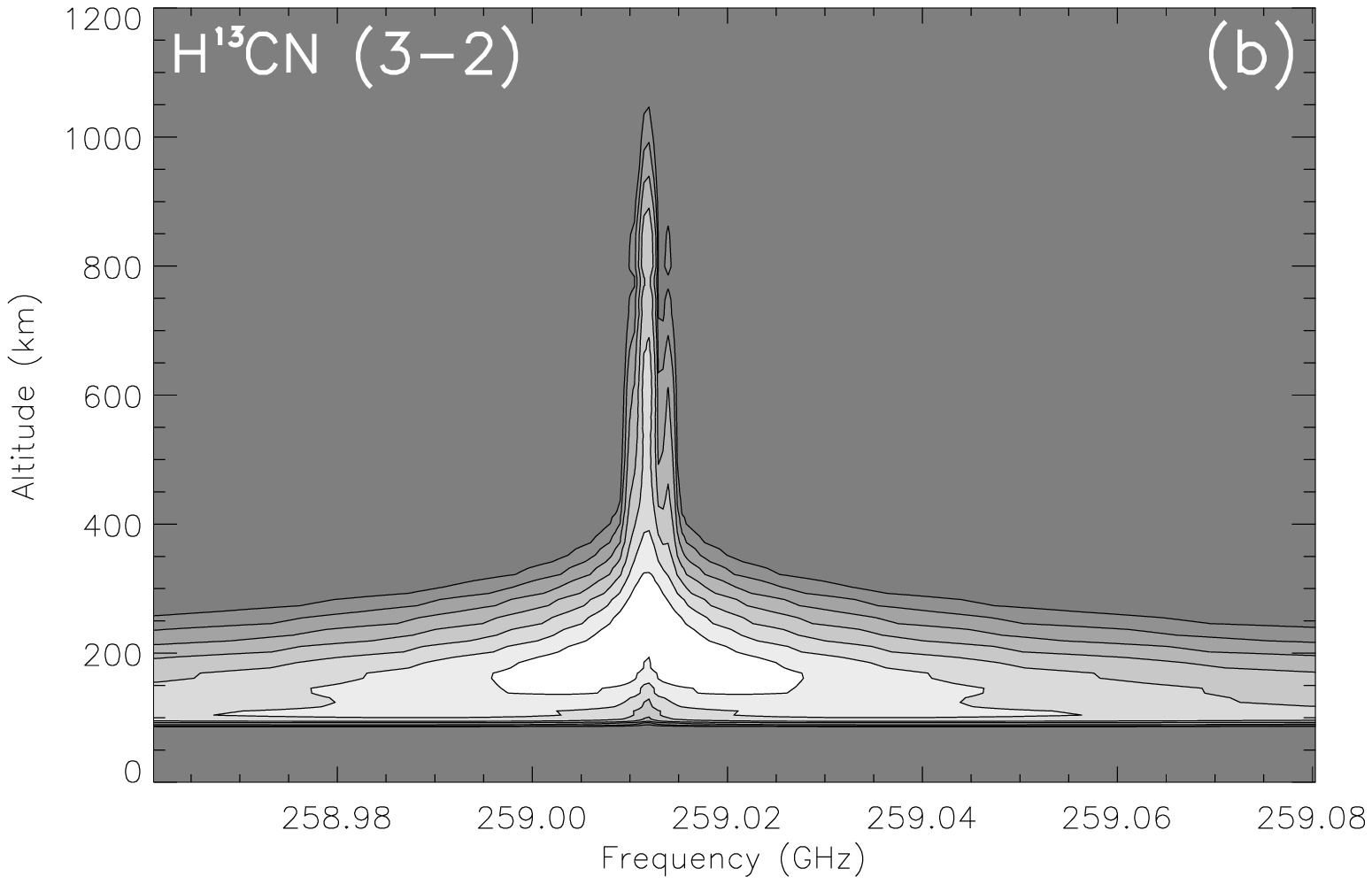}
\includegraphics[width=0.33\textwidth]{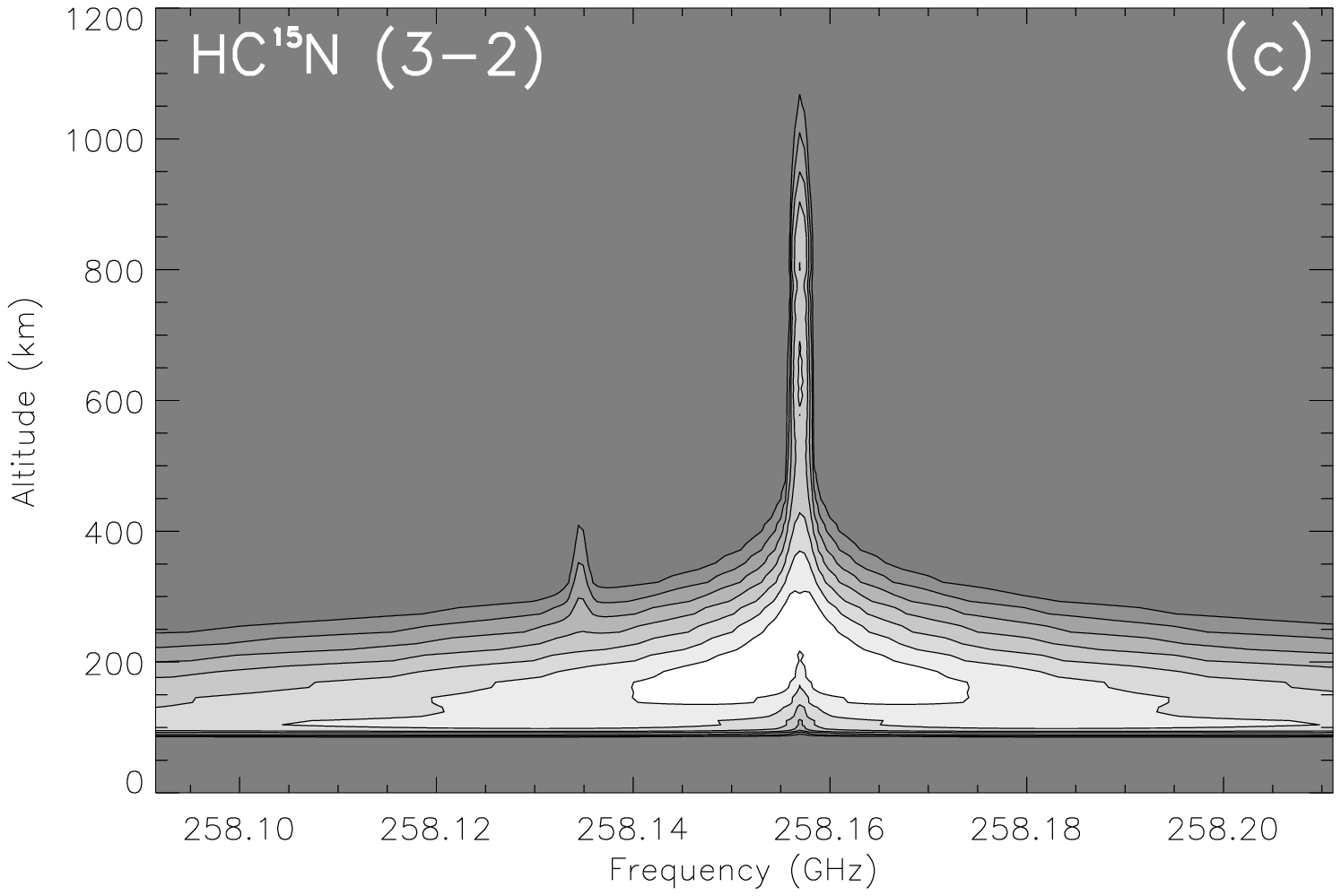} \\
\includegraphics[width=0.33\textwidth]{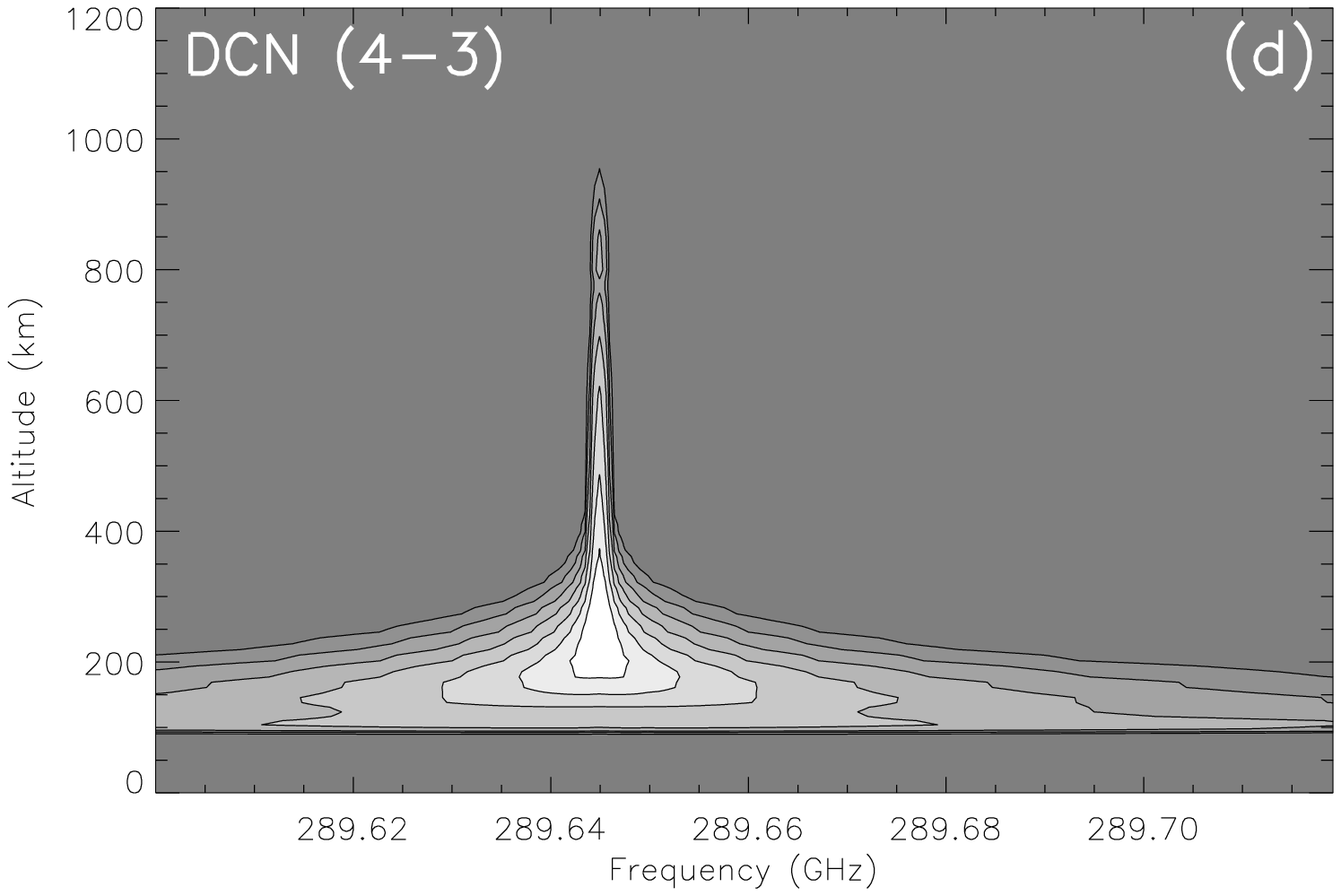}
\includegraphics[width=0.33\textwidth]{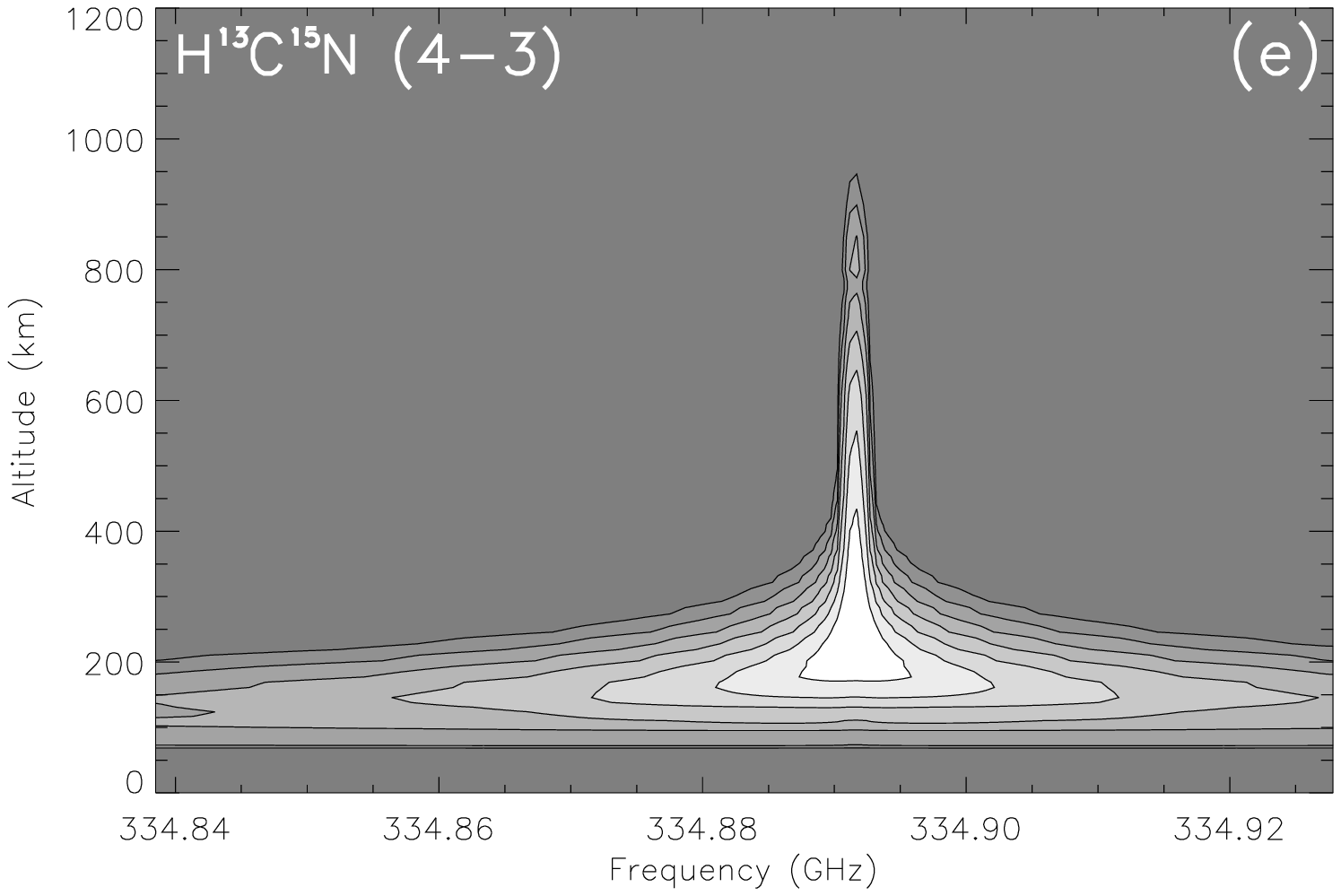}
\includegraphics[width=0.33\textwidth]{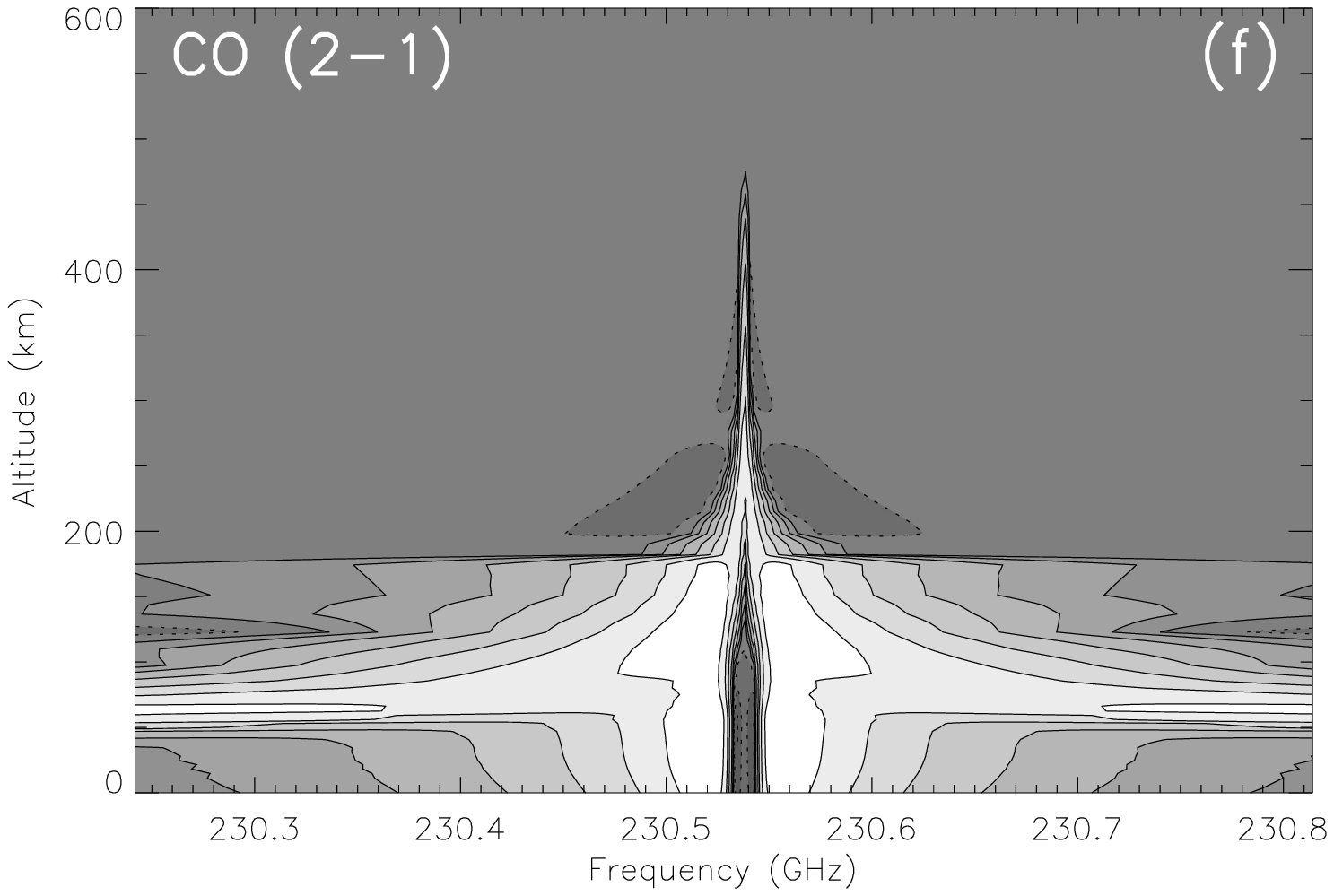} \\
\includegraphics[width=1.0\textwidth]{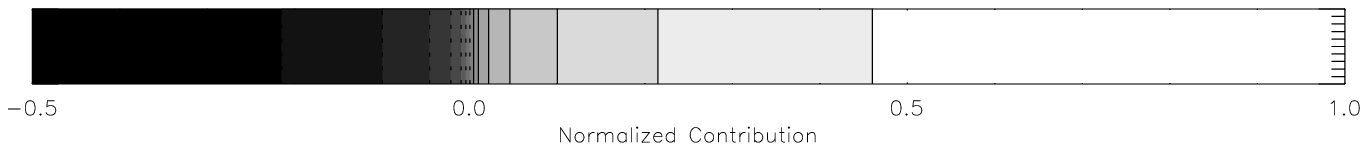}
\caption{Contour plots of the normalized functional derivatives \citep[also called the Jacobians, the matrix of the partial derivatives of radiances at each wavenumber with respect to the retrieved variable; see][]{irwin08} with respect to gas abundance (a-e) or temperature (f) for the model spectra of (a) HCN, (b) \hCn{} (3-2), (c) \hcN{} (3-2), (d) DCN (4-3), (e) \hCN{} (4-3), and (f) CO (2-1). These derivatives depict the altitudes at which the retrieval is sensitive and the variation in sensitivity with wavelength. Contour levels are: 0, $\pm$0.0046, $\pm$0.01, $\pm$0.0215, $\pm$0.046, $\pm$0.1, $\pm$0.215, $\pm$0.46.}
\label{cont_func}
\end{figure*}

We derive a disk-averaged vertical abundance profile from the HCN (1-0) spectral line. NEMESIS uses an iterative $\chi ^2$ minimization technique that relies on both the level of deviation from the a priori setup and the quality of the fit to the data. The error on the a priori profile is taken to be 200\% with a smoothing parameter of three scale heights. We tested a suite of a priori profile error and smoothing values, and found that the chosen values permit the retrieved profile to be constrained primarily by the data while preventing ill-conditioning and unphysical vertical oscillations in the retrieved profile \citep[see discussion in][]{irwin08}. We also determined whether the choice of a priori values affected the retrieved profile by perturbing the a priori abundance profile by two orders of magnitude in each direction; we found that in all cases NEMESIS derived a vertical profile similar to the original best-fit solution (see Figure \ref{hcnprof}b), confirming that the retrieval is well constrained by the data.  We assume the disk-averaged atmospheric temperature profile derived by \citet{serigano16} using an April 2014 observation of the CO (2-1) line, as shown in Figure \ref{hcnprof}a. The fit is sensitive down to $\sim$80 km in the far line wings and up to $\sim$500 km in the line core (see Figure \ref{cont_func}f); below 80 km we allow the profile to relax to the Huygens result and above 500 km an isothermal 160 K atmosphere is assumed in absence of firm temperature constraints. This temperature profile is not allowed to vary, but the temperature errors from the CO line retrieval are carried directly through the matrix inversion within NEMESIS and propagated into the retrieved HCN profile errors.

\begin{figure*}
\includegraphics[angle=270,width=0.33\textwidth]{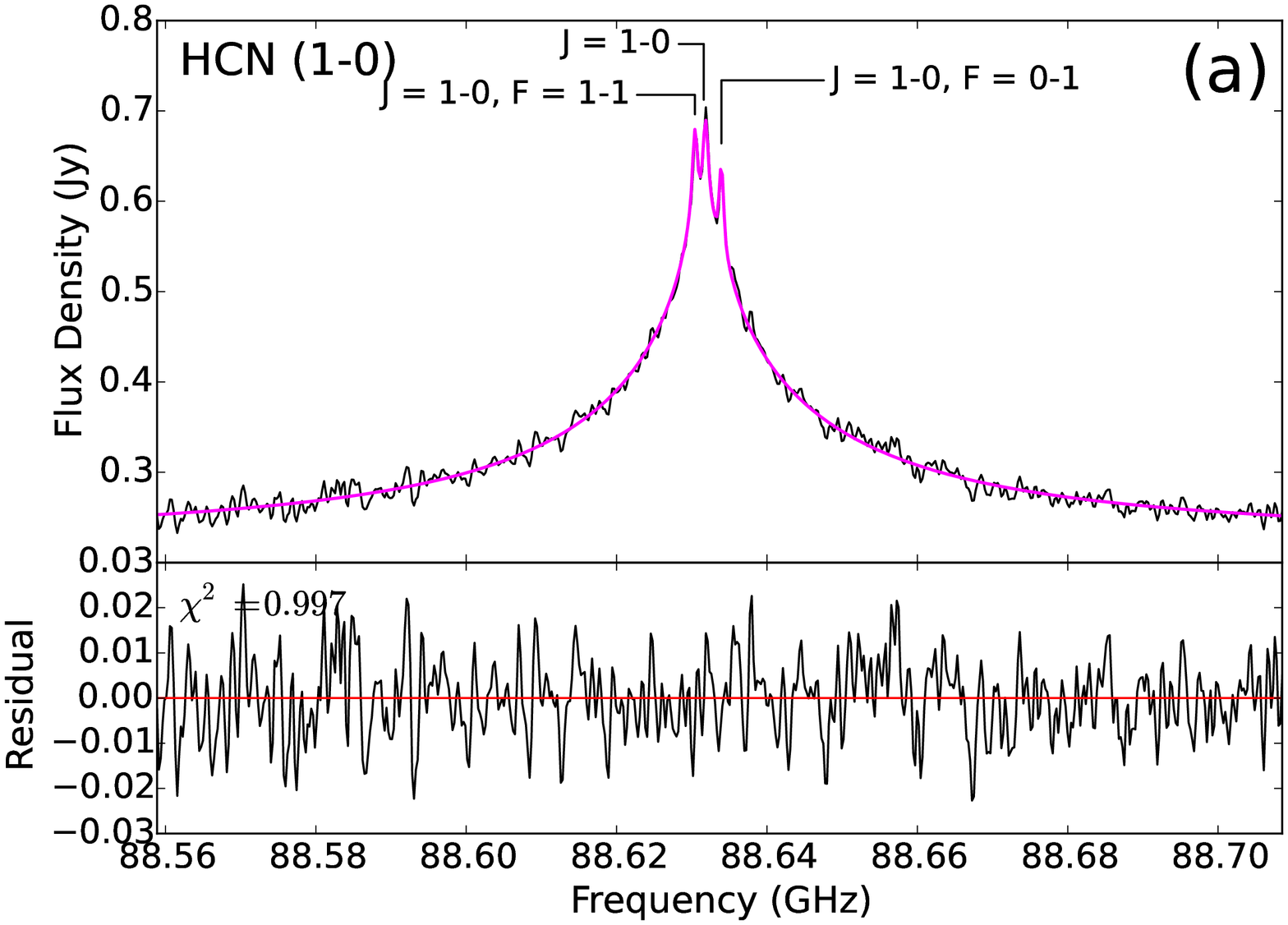}
\includegraphics[angle=270,width=0.33\textwidth]{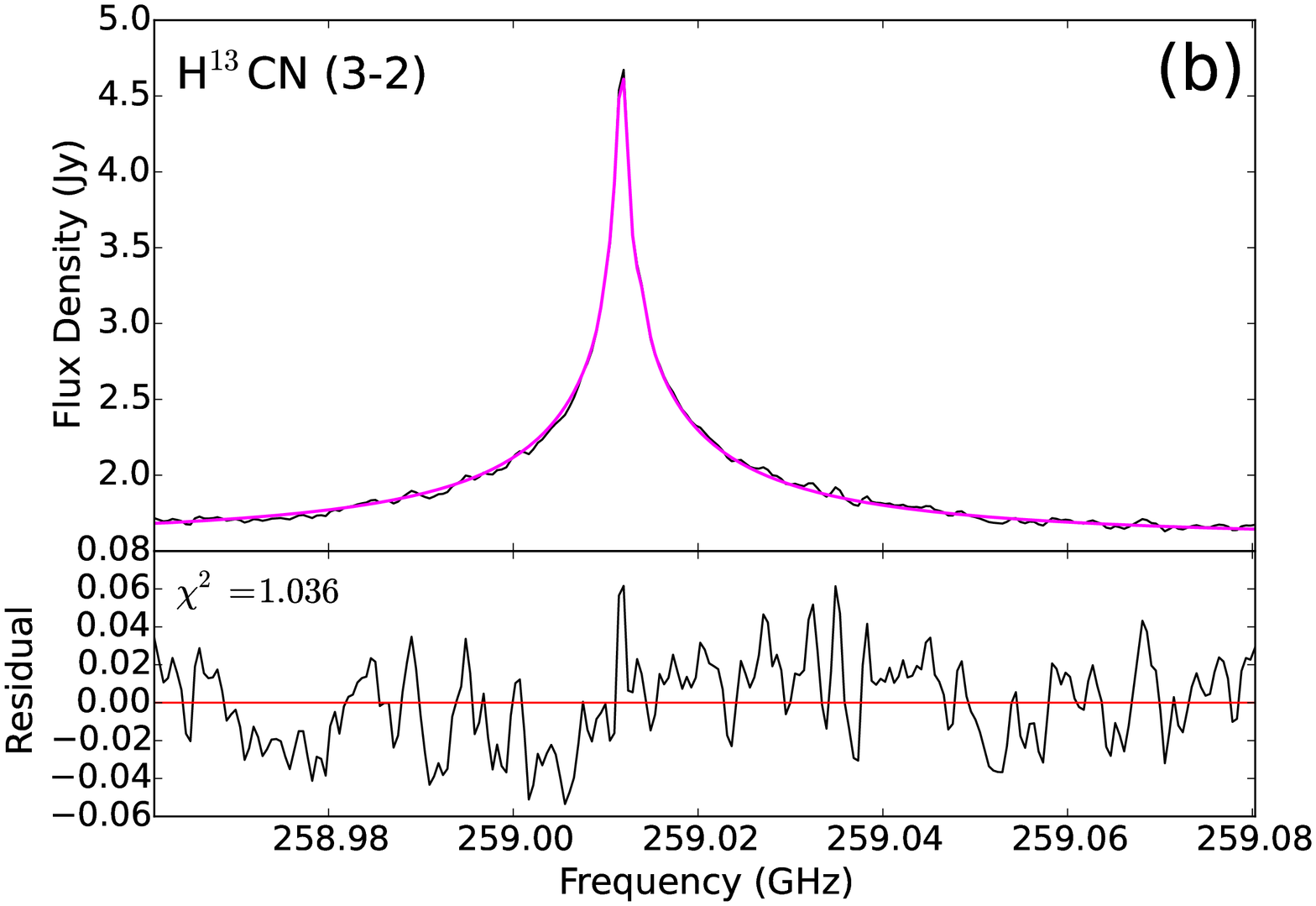}
\includegraphics[angle=270,width=0.33\textwidth]{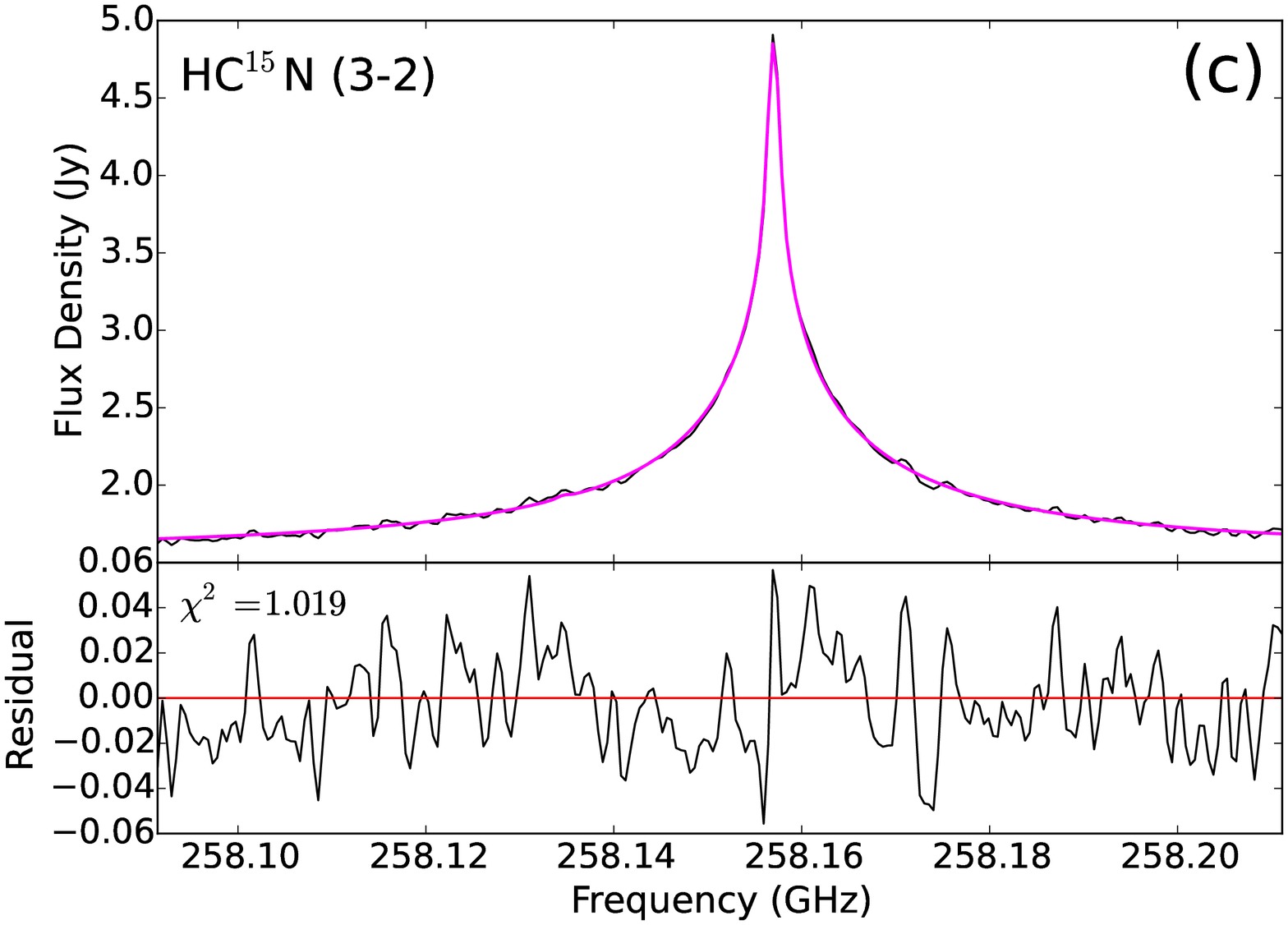}
\includegraphics[angle=270,width=0.33\textwidth]{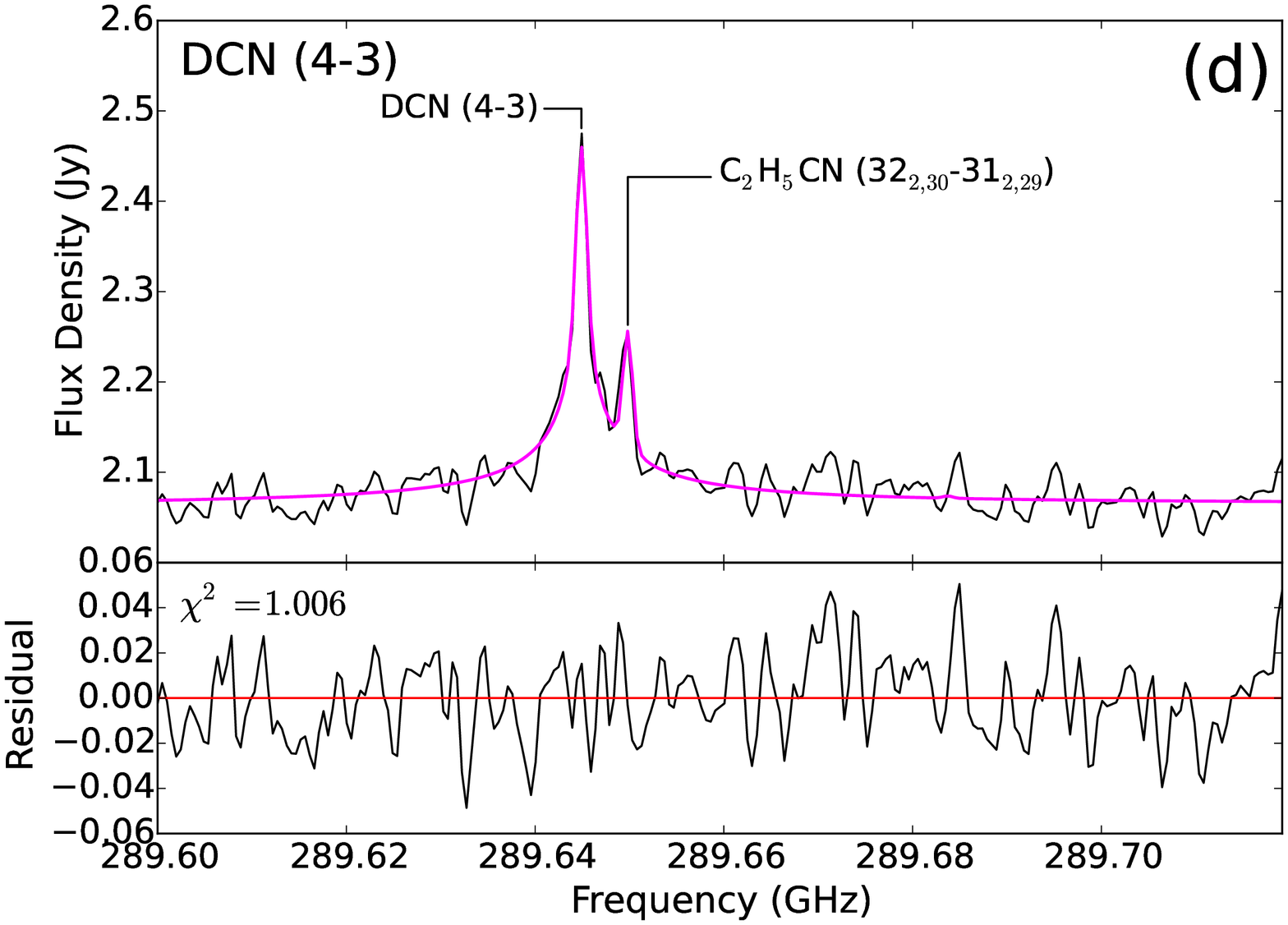}
\includegraphics[angle=270,width=0.33\textwidth]{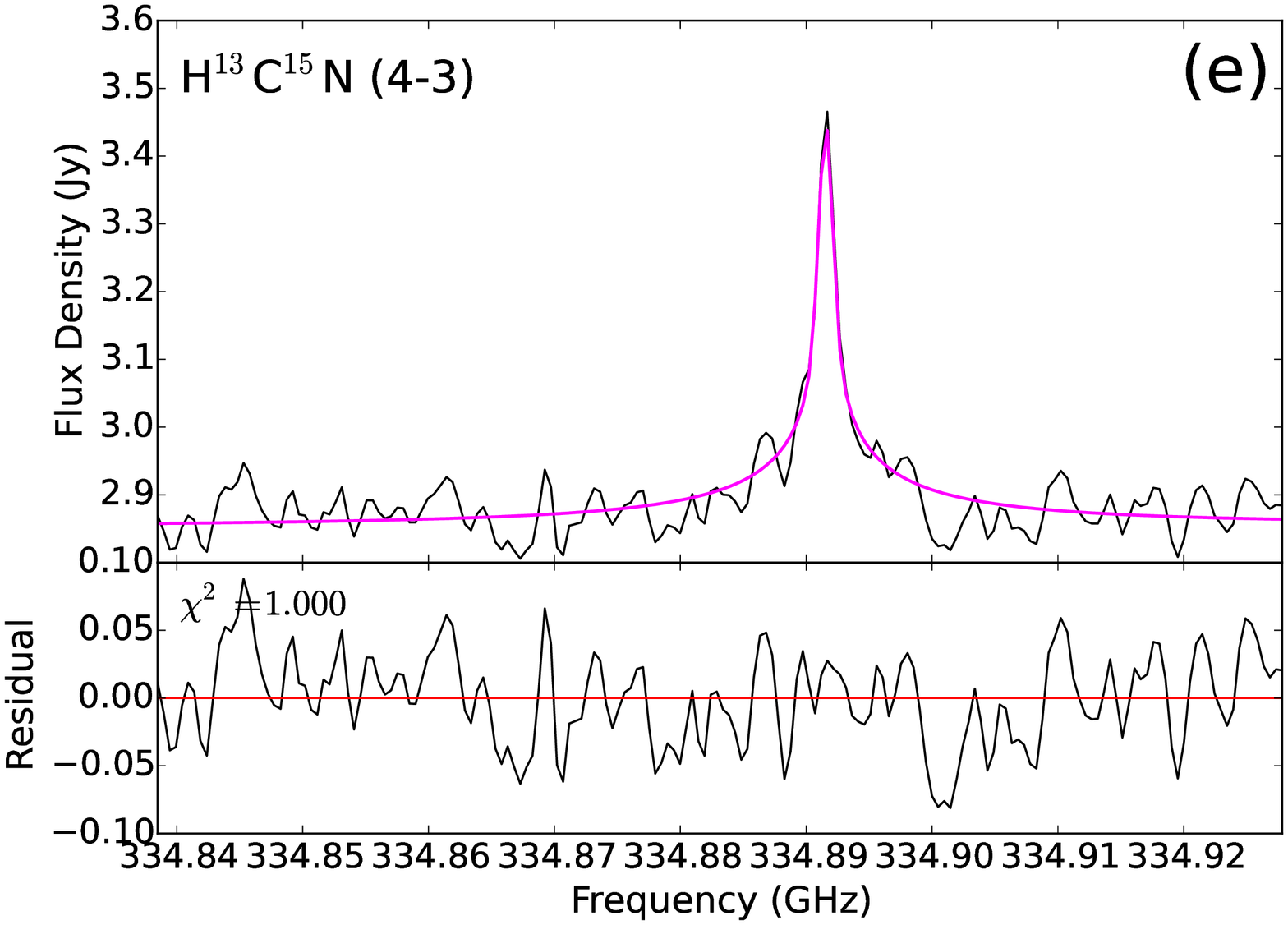}
\includegraphics[angle=270,width=0.33\textwidth]{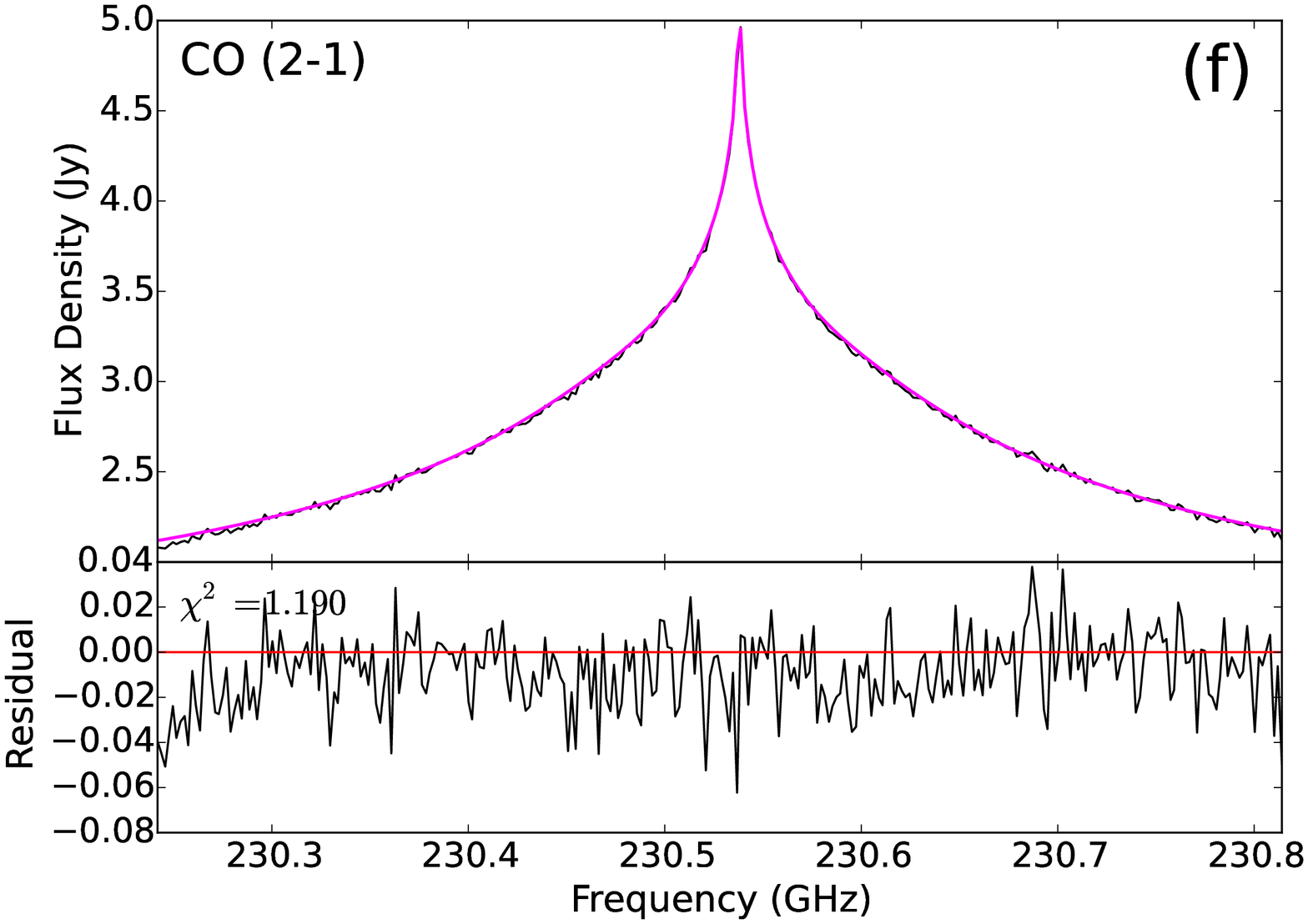}
\caption{Observed (black) and modeled (magenta) spectra for (a) HCN (1-0), (b) \hCn{} (3-2), (c) \hcN{} (3-2), (d) DCN (4-3), (e) \hCN{} (4-3), and (f) CO (2-1). The bottom panel of each subfigure shows the residual flux after subtracting the model from the observed spectrum. The CO fit was performed by \citet{serigano16}. $\chi^2$ is the reduced chi-squared value.}
\label{spectra_hcn}
\end{figure*}

Figure \ref{spectra_hcn}a shows the model fit to the observed HCN (1-0) line assuming the best-fit vertical abundance profile retrieved by NEMESIS, which is shown in Figures \ref{hcnprof}b-d. The fit is sensitive to emission from $\sim$80 km up to $\sim$1100 km (see Figure \ref{cont_func}a); however, the HCN abundance above $\sim$500 km only affects a few data points in the line peaks, and since the temperatures in this region are unconstrained the model has too many free parameters for the derived HCN abundance to be meaningful at these altitudes. That is, for a range of assumed high altitude temperature profiles an abundance profile of HCN that fits the peak structure in the observed spectrum can be found.  We tested many different assumed temperature profiles above 500 km and found that even changes as large as $\pm$50\% had vanishingly small effects on both the VMR profile below 500 km and the derived isotopic ratios.
We assume the same HCN saturation law as \citet{marten02} and \citet{gurwell04}, which forces the gas-phase HCN abundance to zero below $\sim$80 km.

The vertical profiles for the isotopologues were constrained to have the same shape as the vertical profile derived for HCN. NEMESIS was used to retrieve the scaling factor that best fit the observed \hCn{}, \hcN{}, and DCN spectral lines. This best-fit scaling factor corresponds to the ratio between the abundance of the isotopologue and the main species, or isotopic ratio.  We derive the following: \cratio{} = 89.8 $\pm$ 2.8, \nratio{} = 72.3 $\pm$ 2.2, and D/H = (2.5$\pm$0.2)$\times$10$^{-4}$, where the errors correspond to one standard deviation. The \cratio{} and \nratio{} values are a factor of $\sim$3-4 more precise than the most tightly constrained measurements in the literature (see Table \ref{isoRatios}). The D/H measurement is the first published value for a nitrile species on Titan.\footnotemark[8] We model \hCN{} in the same way and find the abundance ratio HCN/\hCN{} = 5800 $\pm$ 270.  The model fits to the data are shown for each of these species in Figure \ref{spectra_hcn}, and the altitudes over which the fits are sensitive are presented in Figure \ref{cont_func}. Since the isotopologue spectra are all well fit by the HCN vertical profile we confirm that assuming a constant isotopic ratio with altitude is acceptable. We note that the HCN (1-0) line appears relatively weak compared to the emission lines from its isotopologues for two reasons: the intrinsic line strength of the HCN (1-0) transition at 150 K is 1-2 orders of magnitude lower than that of the (3-2) and (4-3) transitions, and the very high abundance of HCN leads to saturation of its spectral lines (see Figure \ref{cont_func}).

\footnotetext[8]{A detection of DCN emission has been reported \citep{moreno14}.}

As Figure \ref{spectra_hcn} shows, the ethyl cyanide (32$_{2,30}$-31$_{2,29}$) rovibrational line overlaps with the wing of the DCN (4-3) line. This interloping line is modeled using a 300 km step function for the vertical profile as recommended by \citet{cordiner15}. The 200 km and 400 km step models from that paper were also tested and the choice of model was found to have a negligible effect on the derived DCN abundance.

The total error we report in the derived isotopic ratios combines statistical errors from the RMS noise in the spectrum of the isotopologue, errors in the derived HCN vertical profile (which includes temperature error), errors in the intrinsic line strengths, and errors in the Lorentzian half-width ($\gamma$) and temperature dependence coefficients ($\tau$). The statistical and vertical profile errors are taken into account by NEMESIS directly according to the procedure documented in \citet{irwin08}. We conservatively assume an uncertainty in each of the intrinsic line strengths of 2\% \citep[][as recommended by HITRAN]{maki95}. The error in $\gamma$ and $\tau$ are both estimated to be $\textless$10\% \citep[][as recommended by HITRAN]{devi04,yang08}. We found that varying $\gamma$ by $\pm$10\% changed the derived isotopic ratios by $\lesssim$1\%, and varying $\tau$ by $\pm$10\% affected the ratios at the $\lesssim$0.5\% level. The assumption that $\gamma$ and $\tau$ have the same value for every isotopologue is also imperfect, since increasing mass decreases $\gamma$ according to the definition of the Lorentzian line shape \citep[e.g.,][]{goody89}; however, in this case the difference is only $\sim$1\% between HCN and \hCN{}. We assume all of these errors are uncorrelated and add them in quadrature to our final error estimate. The 3\% continuum rescaling factor leads to an uncertainty in the absolute spectral line intensity; however, this effect is very small compared to the nearly factor-of-ten error ellipse on the derived HCN vertical profile and can be neglected. Since the scaling factor is constant with respect to wavelength its effect on the derived isotopic ratios is also negligible.

Unquantified systematic errors in the derived vertical profile and isotopic ratios may remain.  Systematic errors may arise from assuming that the temperature profile and HCN abundance profile are constant across Titan's disk, that isotopic ratios are constant with altitude, and that the Voigt profile is correct. An accurate estimation of these systematic errors is beyond the scope of this paper.

\section{Discussion}
\label{sec5}

\begin{figure*}
\includegraphics[angle=90, width=0.5\textwidth]{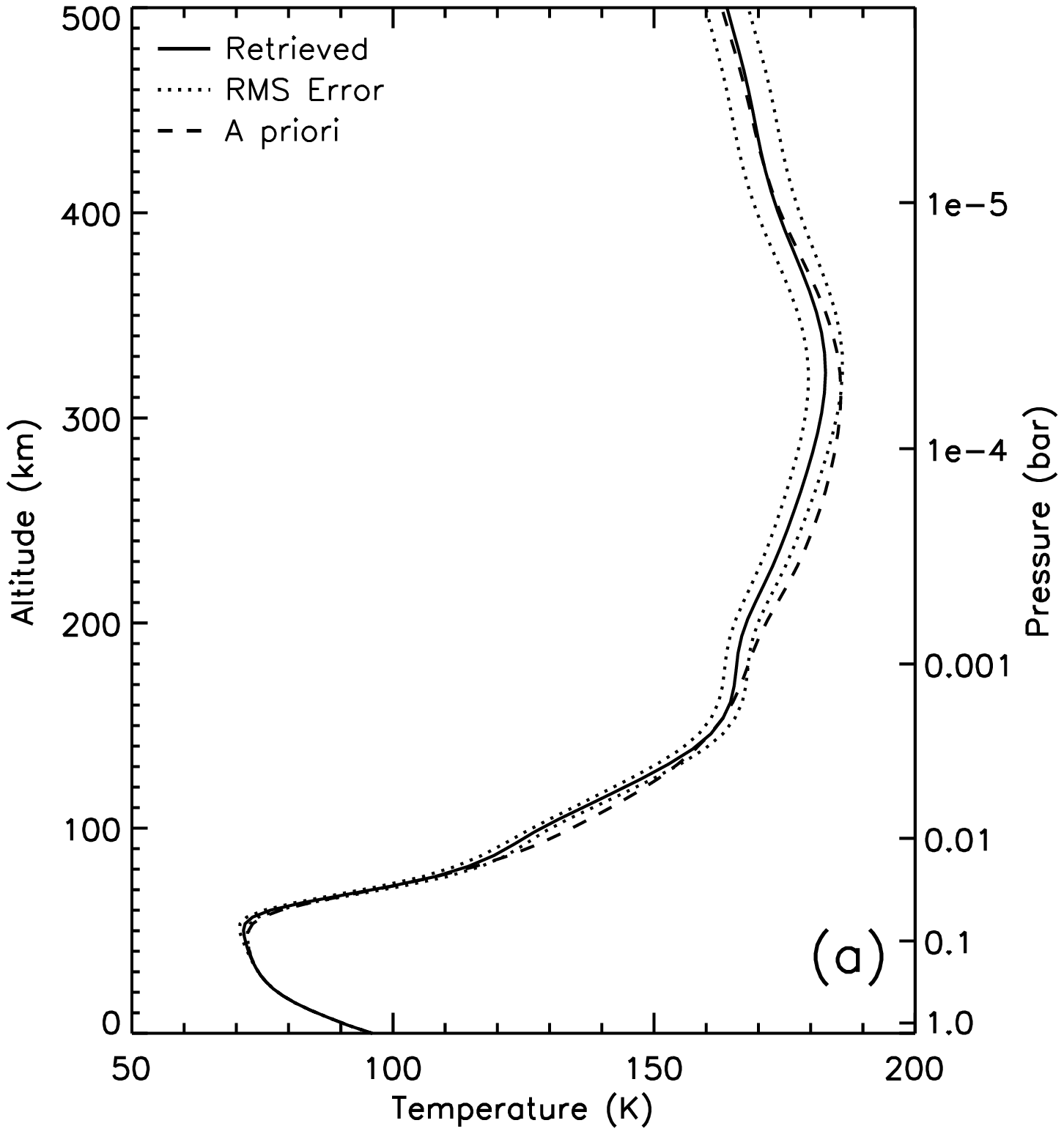}
\includegraphics[angle=90, width=0.5\textwidth]{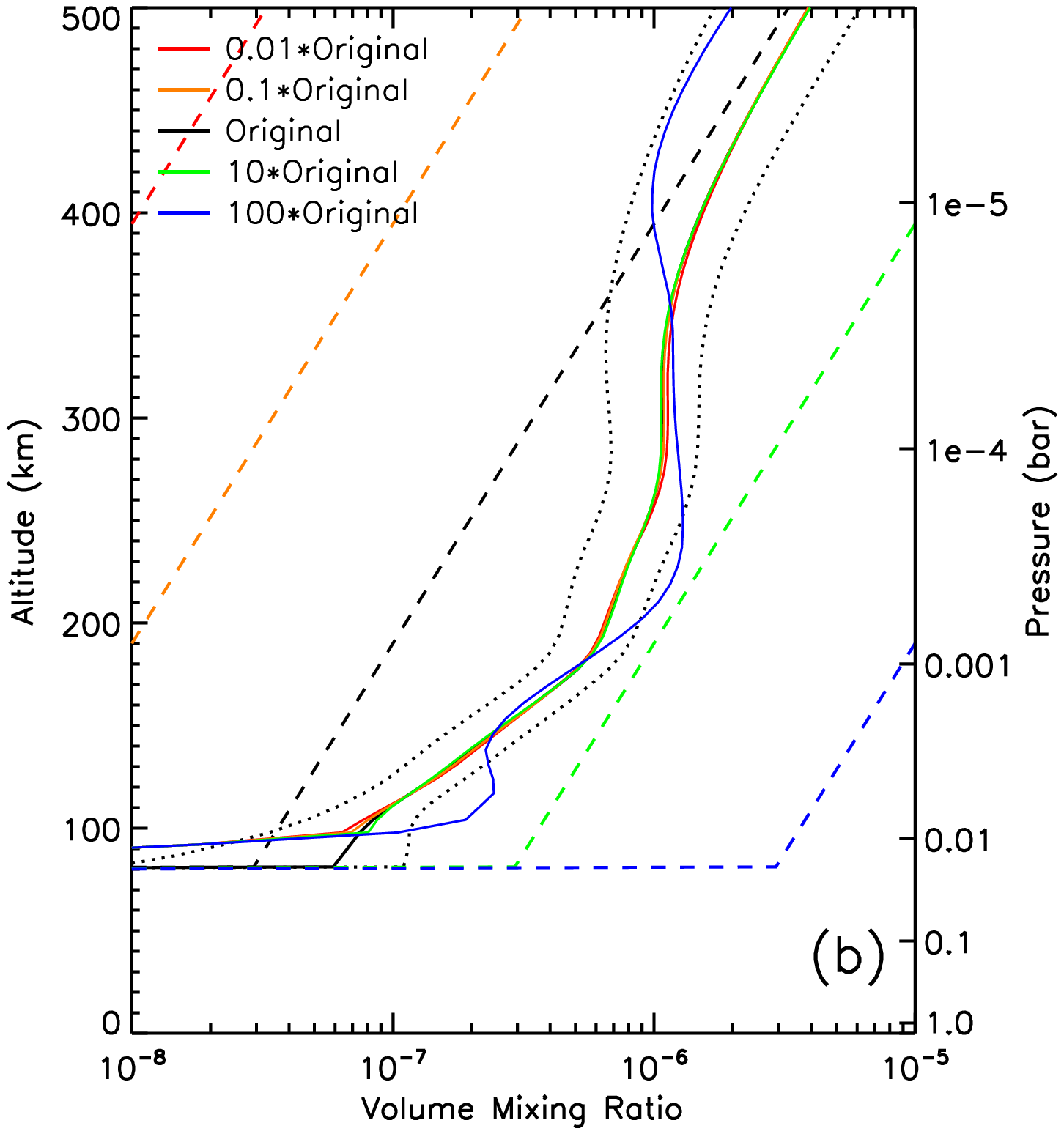}
\includegraphics[angle=90, width=0.5\textwidth]{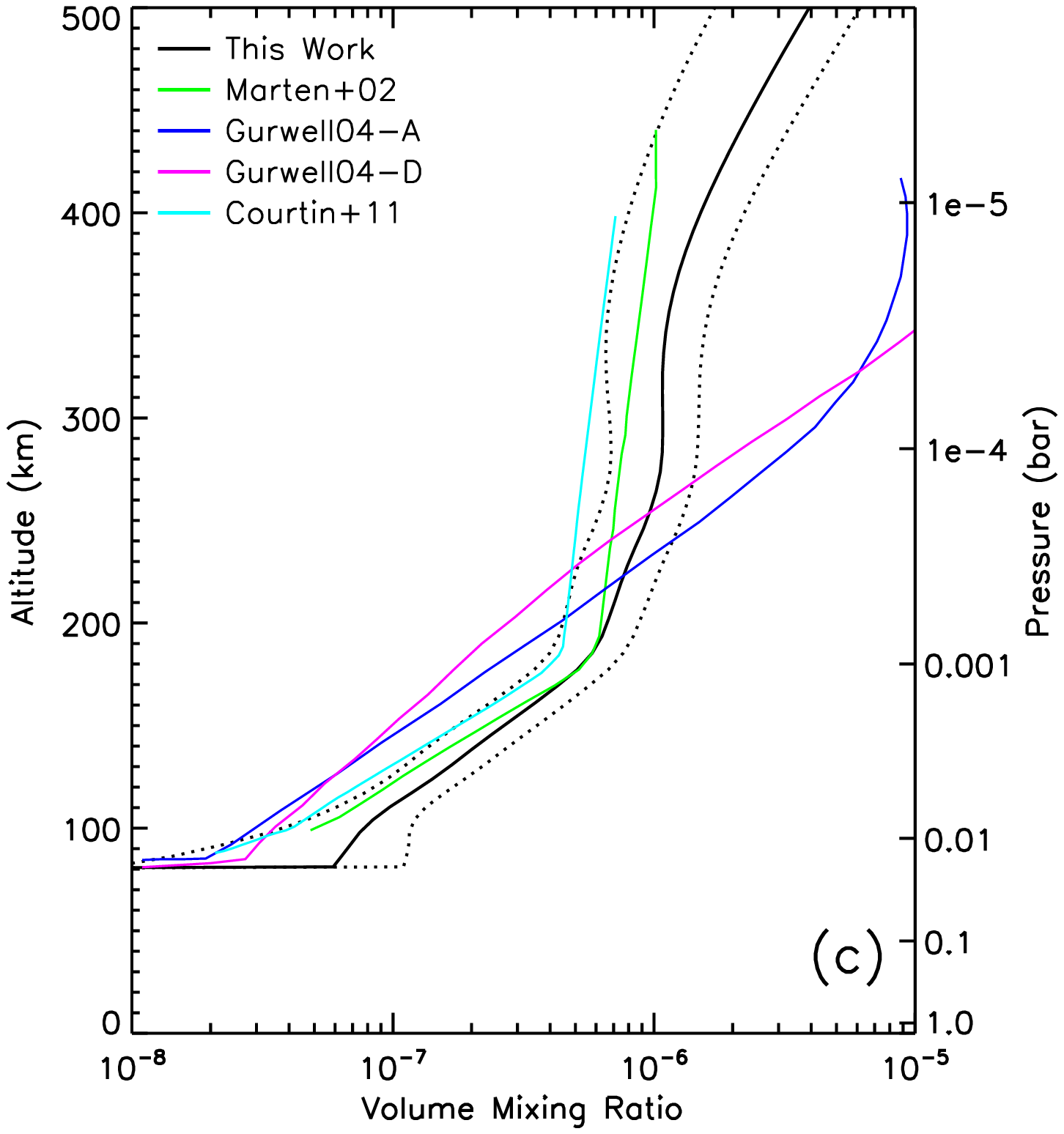}
\includegraphics[angle=90, width=0.5\textwidth]{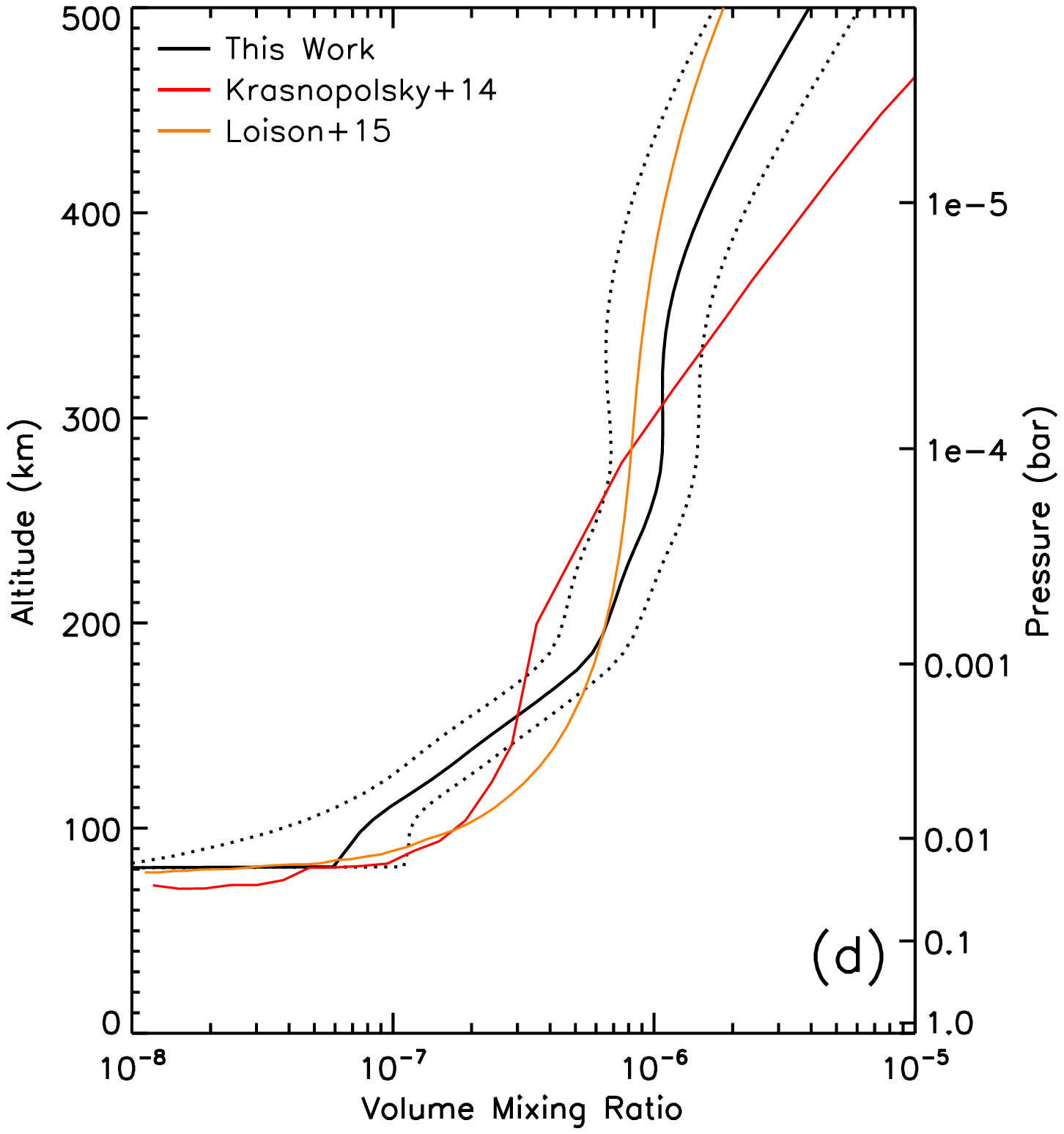}
\caption{(a) Best fit retrieved temperature profile (solid black line) with error band (dotted black lines) and a priori profile (dashed black line) from \citet{serigano16}. (b) Retrieved HCN volume mixing ratio profiles (solid lines) assuming a suite of different a priori profiles (dashed lines) spanning five orders of magnitude. Dotted black lines indicate the RMS error band of the original retrieved profile. (c) Best fit retrieved HCN volume mixing ratio profile compared to observed disk-averaged profiles from the literature \citep[][]{marten02,gurwell04,courtin11}. (d) Comparison of derived HCN vertical profile with predictions from recent photochemical models \citep[][]{krasno14, loison15}.}
\label{hcnprof}
\end{figure*}

The vertical abundance profile we retrieve is compared to similar profiles from the literature in Figure \ref{hcnprof}. The ALMA-derived profile is consistent with those derived from IRAM \citep[][]{marten02} and Herschel \citep[][]{courtin11}, but the abundance increase with altitude is less steep than in the SMA-derived profiles \citep[][]{gurwell04} above $\sim$180 km. We also find fairly strong agreement between our profile and the model profiles put forth by \citet{krasno14} and \citet{loison15}.

The isotopic ratios we report are compared to selected measurements of the \cratio{}, \nratio{}, and D/H ratios from the literature in Table \ref{isoRatios}. \citep[For a recent review of isotopic ratio measurements on Titan see][]{bezard14}. The HCN \cratio{} ratio we report agrees very well with previous infrared, submillimeter, and in situ measurements. The result is also consistent with carbon isotope measurements in CO, CH$_4$, and other hydrocarbons, implying that little to no fractionation of carbon isotopes takes place during the photochemical reactions that produce HCN. Across the solar system the \cratio{} value is found to deviate very little from a single protosolar value of $\sim$89, suggesting a common source for the bulk material \citep[][]{woods09}.

The \nratio{} ratio is found to be a factor of $\sim$2.3 lower than in Titan's N$_2$ \citep[][]{niemann10} and a factor of $\sim$4 lower than the protosolar value \citep[][]{anders89}, as noted by other authors. The photolytic fractionation of N$_2$ is at least partly responsible for this difference; the shift in the rovibrational transition energy of $^{14}$N$^{15}$N makes it self-shield from photodissociation by far-ultraviolet photons less strongly than $^{14}$N$^{14}$N \citep[][]{liang07}, meaning that more atomic $^{15}$N than $^{14}$N is available to produce nitriles in the upper atmosphere. The isotopic ratio reported here is consistent with previous radio observations of Titan but roughly 30\% larger than the value measured by the CIRS instrument on Cassini. One possible source of this mismatch is that the \nratio{} ratio is not independent of altitude; in fact, photochemical models \citep[e.g., ][]{liang07} indicate that this ratio may increase significantly above 750 km due to diffusive separation and other fractionation processes. The CIRS measurement by \citet{vinatier07} was sensitive from 165-305 km while submillimeter observations probe from the condensation altitude ($\sim$80 km) up to at least 450 km \citep[][]{marten02,gurwell04}. Therefore, strong isotopic fractionation as a function of altitude could lead to a systematic difference in the overall isotopic ratio derived using the two techniques. However, the HCN vertical profile derived here fits the observed \hcN{} spectral line down to the RMS noise level, so we conclude that these data do not provide evidence for fractionation.

The D/H ratio in hydrogen and methane on Titan is known to be significantly elevated compared to the protosolar value, providing important constraints on photochemical enrichment, mass-dependent escape, and perhaps a primordial deuterium enrichment in Titan's atmosphere \citep[e.g.,][]{cordier08}.  In Table \ref{isoRatios} the ALMA measurement of the D/H ratio in HCN on Titan is compared to literature measurements of D/H in H$_2$, CH$_4$, and C$_2$H$_2$ (acetylene). The value we report is elevated by a factor of $\sim$2 compared to the D/H ratio found in molecular hydrogen and methane on Titan by Cassini infrared measurements but consistent with the ratio found in acetylene, implying further enrichment in deuterium taking place during one or more of the chemical reactions that form hydrocarbons and nitriles in Titan's atmosphere. The kinetic isotope effect may be responsible for this discrepancy: the C-H bond is more easily photolysed than the C-D bond in methane \citep[][]{bezard14}, causing more H than D to escape into space and more CH$_2$D than CH$_3$ to participate in the chemical reactions that create HCN and C$_2$H$_2$. In addition, hydrodynamic escape is more rapid for hydrogen atoms than deuterium atoms, leading to a net enrichment in deuterium in Titan's photochemistry. Our measurement thus helps to constrain the photochemical and mass-dependent fractionation processes on Titan, but a detailed analysis of these is beyond the scope of this paper.

Following the decommissioning of Cassini in September 2017, further study of the dynamics and evolution of Titan's atmosphere will rely on ground- and space-based observatories. The observations in this paper demonstrate the immense potential of ALMA to expand upon the advances made by Cassini. As more antennas come online, longer baselines are utilized, and dedicated hours-long observations are carried out, ALMA will become an indispensable tool for mapping latitudinal and longitudinal distributions of molecules, tracking seasonal changes, and searching for new photochemical products both on Titan and elsewhere in the Solar System.

\acknowledgements

\noindent This research was supported by NASA's Planetary Atmospheres and Planetary Astronomy programs.

\noindent JEL is supported by an appointment to the NASA Postdoctoral Program at the NASA Goddard Space Flight Center, administered by Universities Space Research Association through a contract with NASA.

\noindent We thank the staff at the helpdesk of the North American ALMA Science Center (NAASC) in Charlottesville, Virginia for providing helpful information on the handling of ALMA data.

\noindent This paper makes use of the following ALMA datasets: ADS/JAO.ALMA \#2012.1.00261.S, \#2012.1.00453.S and \#2012.1.00566.S. ALMA is a partnership of ESO (representing its member states), NSF (USA) and NINS (Japan), together with NRC (Canada) and NSC and ASIAA (Taiwan) and KASI (Republic of Korea), in cooperation with the Republic of Chile. The Joint ALMA Observatory is operated by ESO, AUI/NRAO and NAOJ. The National Radio Astronomy Observatory is a facility of the National Science Foundation operated under cooperative agreement by Associated Universities, Inc.


\clearpage
\begin{landscape}
\begin{deluxetable}{lccccccccc}
\tablecaption{Observational Parameters}  
\tablewidth{0pt}  
\tablehead{
\colhead{Species \&} &\colhead{Rest Freq.} &\colhead{Obs.} &\colhead{Integration} &\colhead{No. of} &\colhead{Spectral} &\colhead{Beam} &\colhead{Distance} &\colhead{Velocity} &\colhead{Project} \\
\colhead{Transition} &\colhead{(GHz)} &\colhead{Date} &\colhead{Time (s)} &\colhead{Antennas} &\colhead{Res. (kHz)\tablenotemark{a}} &\colhead{Size (\arcsec)\tablenotemark{b}} &\colhead{(AU)\tablenotemark{c}} &\colhead{(\kms{})\tablenotemark{c}} &\colhead{ID}}
\startdata

HCN (1-0) &88.631 &2014-04-03 &157 &32 &488 &2.48 $\times$ 1.89 &9.10762 &-23.486 &2012.1.00566.S\\

\hCn{} (3-2) &259.012 &2014-07-07 &157 &31 &976 &0.44 $\times$ 0.41 &9.36015 &19.268 &2012.1.00453.S\\

\hcN{} (3-2) &258.157 &2014-07-07 &157 &31 &976 &0.44 $\times$ 0.41 &9.36015 &19.268 &2012.1.00453.S\\

DCN (4-3) &362.045 &2014-06-16 &157 &35 &976 &0.49 $\times$ 0.41 &9.09498 &21.495 &2012.1.00453.S\\

\hCN{} (4-3) &334.891 &2014-05-27 &158 &31 &976 &0.46 $\times$ 0.39 &8.93378 &11.785 &2012.1.00453.S \\

CO (2-1) &230.538 &2014-04-04 &158 &34 &1953 &0.87 $\times$ 0.72 &9.09365 &-23.121 &2012.1.00261.S\\ 

\enddata
\tablenotetext{a}{After channel smoothing by the correlator; twice the channel spacing.}
\tablenotetext{b}{Full width at half-maximum of the Gaussian restoring beam.}
\tablenotetext{c}{Radial distance and velocity with respect to observer as calculated by JPL Horizons (\texttt{http://ssd.jpl.nasa.gov/horizons.cgi}).}
\label{obsDetails}
\end{deluxetable}
\end{landscape}

\clearpage
\begin{deluxetable}{ccccc}
\tablecaption{Recent Measurements of Isotopic Ratios}  
\tablewidth{0pt}  
\tablehead{
\colhead{Ratio} &\colhead{Measurement} &\colhead{Species} &\colhead{Instrument/Waveband} &\colhead{Reference}}
\startdata

\cratio{} &91.1$\pm$1.4 &CH$_4$ &Huygens GCMS &\citet{niemann10}\\
&86.5$\pm$7.9 & &Cassini CIRS/IR &\citet{nixon12}\\
&89.9$\pm$3.4 &CO &ALMA/(sub)mm &\citet{serigano16}\\
&108$\pm$20 &HCN &SMA/(sub)mm &\citet{gurwell04} A\\
&132$\pm$25 & &SMA/(sub)mm &\citet{gurwell04} D\\
&79$\pm$17 & &Cassini CIRS/IR &\citet{vinatier07}\\
&96$\pm$13 & &Herschel SPIRE/(sub)mm &\citet{courtin11}\\
&66$\pm$35 & &Herschel PACS/(sub)mm &\citet{rengel14}\\
&89.8$\pm$2.8 & &ALMA/(sub)mm &This Work\\
\\
\hline
\\
\nratio{} &167$\pm$0.6 &N$_2$ &Huygens GCMS &\citet{niemann10}\\
&65$\pm$6.5 &HCN &IRAM/submm &\citet{marten02}\\
&72$\pm$9 & &SMA/submm &\citet{gurwell04} A\\
&94$\pm$13 & &SMA/submm &\citet{gurwell04} D\\
&56$\pm$8 & &Cassini CIRS/IR &\citet{vinatier07}\\
&65$\pm$12 & &SMA/submm &\citet{gurwell11}\\
&76$\pm$6 & &Herschel SPIRE/submm &\citet{courtin11}\\
&72.2$\pm$2.2 & &ALMA/submm &This Work\\
\\
\hline
\\
D/H &(1.35$\pm$0.30)$\times$10$^{-4}$ &H$_2$ &Huygens GCMS &\citet{niemann10}\\
&(1.32$^{+0.15}_{-0.11} \times$10$^{-4}$ &CH$_4$ &Cassini CIRS/IR &\citet{bezard07} \\
&(1.59$\pm$0.27)$\times$10$^{-4}$ &CH$_4$ &Cassini CIRS/IR &\citet{nixon12}\\ 
&(2.09$\pm$0.45)$\times$10$^{-4}$ &C$_2$H$_2$ &Cassini CIRS/IR &\citet{coustenis08}\\
&(2.5$\pm$0.2)$\times$10$^{-4}$ &HCN &ALMA/submm &This Work\\

\enddata
\label{isoRatios}
\end{deluxetable}

\clearpage

\clearpage

\clearpage

\clearpage
\bibliographystyle{apj}                                                 

\begin{thebibliography}{}


\bibitem[Ahrens et al.(2002)]{ahrens02} Ahrens, V., Lewen, F., 
Takano, S., et al.\ 2002, Zeitschrift Naturforschung Teil A, 57, 669 

\bibitem[Anders 
\& Grevesse(1989)]{anders89} Anders, E., \& Grevesse, N.\ 1989, \gca, 53, 197 


\bibitem[B{\'e}zard et al.(2007)]{bezard07} B{\'e}zard, B., 
Nixon, C.~A., Kleiner, I., \& Jennings, D.~E.\ 2007, \icarus, 191, 397 

\bibitem[B\'{e}zard et al.(2014)]{bezard14} B\'{e}zard, B., Yelle R. \& Nixon C.~A.\ 2014, Titan: Surface, Atmosphere and Magnetosphere ed. C., Muller-Wodarg et al. (Cambridge: Cambridge Univ. Press) 158 

\bibitem[Br{\"u}nken et al.(2004)]{brunken04} Br{\"u}nken, S., 
Fuchs, U., Lewen, F., et al.\ 2004, Journal of Molecular Spectroscopy, 225, 
152  

\bibitem[Cordier et al.(2008)]{cordier08} Cordier, D., Mousis, 
O., Lunine, J.~I., Moudens, A., \& Vuitton, V.\ 2008, \apjl, 689, L61

\bibitem[Cordiner et al.(2014)]{cordiner14} Cordiner, M.~A., 
Nixon, C.~A., Teanby, N.~A., et al.\ 2014, \apjl, 795, L30 

\bibitem[Cordiner et al.(2015)]{cordiner15} Cordiner, M.~A., 
Palmer, M.~Y., Nixon, C.~A., et al.\ 2015, \apjl, 800, L14

\bibitem[Courtin et 
al.(2011)]{courtin11} Courtin, R., Swinyard, B.~M., Moreno, R., et al.\ 2011, \aap, 536, L2 

\bibitem[Coustenis et al.(1991)]{coustenis91} Coustenis, A., 
Bezard, B., Gautier, D., Marten, A., 
\& Samuelson, R.\ 1991, \icarus, 89, 152 


\bibitem[Coustenis et al.(2008)]{coustenis08} Coustenis, A., 
Jennings, D.~E., Jolly, A., et al.\ 2008, \icarus, 197, 539 


\bibitem[Devi et al.(2004)]{devi04} Devi, V.~M., Benner, 
D.~C., Smith, M.~A.~H., et al.\ 2004, \jqsrt, 87, 339 



\bibitem[Fuchs et al.(2004)]{fuchs04} Fuchs, U., Bruenken, S., 
Fuchs, G.~W., et al.\ 2004, Zeitschrift Naturforschung Teil A, 59, 861 



\bibitem[Goody \& Yung(1989)]{goody89} Goody, R.~M., \& Yung, Y.~L.\ 1989, Atmospheric radiation : theoretical basis, 2nd ed., by Richard M.~Goody and Y.L.~Yung.~ New York, NY: Oxford University Press, 1989,  

\bibitem[Gurwell(2004)]{gurwell04} Gurwell, M.~A.\ 2004, \apjl, 
616, L7

\bibitem[Gurwell et al.(2011)]{gurwell11} Gurwell, M., Moreno, 
R., Moullet, A., \& Butler, B.\ 2011, EPSC-DPS Joint Meeting 2011, 270 

\bibitem[Hidayat et al.(1997)]{hidayat97} Hidayat, T., Marten, 
A., B{\'e}zard, B., et al.\ 1997, \icarus, 126, 170 

\bibitem[Irwin et al.(2008)]{irwin08} Irwin, P.~G.~J., Teanby, 
N.~A., de Kok, R., et al.\ 2008, \jqsrt, 109, 1136




\bibitem[Khare et al.(1986)]{khare86} Khare, B.~N., Sagan, C., 
Ogino, H., et al.\ 1986, \icarus, 68, 176

\bibitem[Koskinen et al.(2011)]{koskinen11} Koskinen, T.~T., 
Yelle, R.~V., Snowden, D.~S., et al.\ 2011, \icarus, 216, 507

\bibitem[Krasnopolsky(2014)]{krasno14} Krasnopolsky, V.~A.\ 
2014, \icarus, 236, 83


\bibitem[Liang et al.(2007)]{liang07} Liang, M.~C., Heays, 
A.~N., Lewis, B.~R., Gibson, S.~T., \& Yung, Y.~L.\ 2007, \apjl, 664, L115

\bibitem[Loison et al.(2015)]{loison15} Loison, J.~C., 
H{\'e}brard, E., Dobrijevic, M., et al.\ 2015, \icarus, 247, 218

\bibitem[Maki et al.(1995)]{maki95} Maki, A., Quapp, W., Klee, S., Mellau, G.~C., \& Albert, S.\ 1995, Journal of Molecular Spectroscopy, 174, 365 

\bibitem[Marten et al.(2002)]{marten02} Marten, A., Hidayat, T., 
Biraud, Y., \& Moreno, R.\ 2002, \icarus, 158, 532

\bibitem[Moreno et al.(2014)]{moreno14} Moreno, R., Lellouch, 
E., Vinatier, S., et al.\ 2014, AAS/Division for Planetary Sciences Meeting 
Abstracts, 46, \#211.19 

\bibitem[M{\"u}ller et 
al.(2001)]{muller01} M{\"u}ller, H.~S.~P., Thorwirth, S., Roth, D.~A., \& Winnewisser, G.\ 2001, \aap, 370, L49 

\bibitem[Niemann et al.(2010)]{niemann10} Niemann, H.~B., Atreya, 
S.~K., Demick, J.~E., et al.\ 2010, Journal of Geophysical Research 
(Planets), 115, E12006 

\bibitem[Nixon et al.(2012)]{nixon12} Nixon, C.~A., Temelso, 
B., Vinatier, S., et al.\ 2012, \apj, 749, 159 



\bibitem[Rengel et al.(2014)]{rengel14} Rengel, M., Sagawa, H., Hartogh, P., et al.\ 2014, \aap, 561, A4 

\bibitem[Rothman et al.(2005)]{rothman05} Rothman, L.~S., 
Jacquemart, D., Barbe, A., et al.\ 2005, \jqsrt, 96, 139 

\bibitem[Sagan et al.(1992)]{sagan92} Sagan, C., Thompson, W.~R., \& Khare, B.~N.\ 1992, Accounts of Chemical Research, 25.7

\bibitem[Serigano et al.(2016)]{serigano16} Serigano, J., Nixon, C.~A., Cordiner, M.~A., et al.\ 2016, \apjl, 821, L8 

\bibitem[Stevenson et al.(2015)]{stevenson15} Stevenson, J., 
Lunine, J., \& Clancy, P.\ 2015, Science Advances, 1, 1400067 

\bibitem[Stofan et al.(2007)]{stofan07} Stofan, E.~R., Elachi, 
C., Lunine, J.~I., et al.\ 2007, \nat, 445, 61

\bibitem[Teanby et al.(2007)]{teanby07} Teanby, N.~A., Irwin, 
P.~G.~J., de Kok, R., et al.\ 2007, \icarus, 186, 364

\bibitem[Teanby et al.(2010)]{teanby10} Teanby, N.~A., Irwin, 
P.~G.~J., de Kok, R., \& Nixon, C.~A.\ 2010, Faraday Discussions, 147, 51 

\bibitem[Teanby et 
al.(2013)]{teanby13} Teanby, N.~A., Irwin, P.~G.~J., Nixon, C.~A., et al.\ 2013, \planss, 75, 136 

\bibitem[Vinatier et al.(2007)]{vinatier07} Vinatier, S., 
B{\'e}zard, B., \& Nixon, C.~A.\ 2007, \icarus, 191, 712

\bibitem[Vinatier et al.(2010)]{vinatier10} Vinatier, S., 
B{\'e}zard, B., Nixon, C.~A., et al.\ 2010, \icarus, 205, 559 


\bibitem[Wilson 
\& Atreya(2004)]{wilson04} Wilson, E.~H., \& Atreya, S.~K.\ 2004, Journal of Geophysical Research (Planets), 109, E06002 

\bibitem[Woods(2009)]{woods09} Woods, P.~M.\ 2009, 
arXiv:0901.4513  

\bibitem[Yang et al.(2008)]{yang08} Yang, C., Buldyreva, J., 
Gordon, I.~E., et al.\ 2008, \jqsrt, 109, 2857 

\end{thebibliography}


\end{document}